\documentclass[twocolumn,twoside,preprintnumbers,amsmath,amssymb,showkeys]{revtex4}
\usepackage{epsfig}
\usepackage{graphicx}

\usepackage{fancyhdr}

\usepackage{pslatex}

\begin{document}

\title{Fluctuations, correlations and non-extensivity}

\author{Grzegorz Wilk}
\affiliation{The Andrzej So\l tan Institute for Nuclear Studies, Ho\.za
69, 00681 Warsaw, Poland}


\begin{abstract}
The present status of investigations on fluctuations and correlations
seen in high energy multiparticle production processes made using the
notion of nonextensivity is reviewed.


\end{abstract}
\maketitle

\thispagestyle{fancy}

\setcounter{page}{1}

\section{Introduction}

There are many examples that distributions of produced particles deviate
from the expected exponential form towards its power-like generalization
\cite{examples}, called the $q$-exponential and know also as Tsallis
distribution \cite{CT1},
\begin{equation}
\exp_q\left(-\frac{X}{\lambda}\right)\stackrel{def}{=}\left[1-(1-q)\frac{X}{\lambda}
\right]^{1/(1-q)}\stackrel{q\rightarrow 1}{\Longrightarrow} \exp \left(
-\frac{X}{\lambda} \right), \label{eq:defq}
\end{equation}
described by parameter $q$ (in cases of interest to us $X=p_T$ or
$X=m_T\cosh y$). Distributions of this type were proposed on
phenomenological grounds long time ago as useful parametrization of data
interpolating between results coming from {\it soft} and {\it hard}
scattering \cite{CM}. On the other hand, when seen from the
information-theory point of view or from the thermodynamical approach,
such distributions arise from the use of non-extensive Tsallis entropy
\cite{T} (indexed by $q$) instead of the usual Shannon-Gibbs-Boltzmann
one:
\begin{equation}
S_q = -\, \frac{1}{1 - q} \left( 1\, -\, \sum_i p_i^q \right)
        \stackrel{q \rightarrow 1}{\Longrightarrow}
        S_{SBG}\, =\, -\, \sum_i p_i\, \ln p_i .
 \label{eq:Def}
\end{equation}
Notice that $S_q$ is nonextensive in such sense that for  two independent
systems, $A$ and $B$ (in the usual sense, i.e., that the corresponding
probabilities of their occurrence factorize, $p_{ij}(A+B)=p_i(A)p_j(B)$)
one has that
\begin{equation}
S_q(A+B)\, = \, S_q(A)\, +\, S_q(B)\, + (1-q)\, S_q(A)S_q(B) .
         \label{eq:Nonex}
\end{equation}
The entropic index $q$ is thus also a measure of nonextensivity in system
under consideration and therefore is also called non-extensivity
parameter.

What are the circumstances leading to $q\neq 1$? The common consensus is
that it happens whenever \cite{T}: - there are {\it long range
correlations} in the system (or {\it system is small}~-~ notice that our
Universe is small with respect to the gravitational interactions!); -
there are {\it memory effects} of different kind; - the phase-space in
which system operates is limited or has {\it fractal structure} and,
finally, - there are {\it intrinsic fluctuations} in the system under
consideration.

\section{The role of intrinsic fluctuations}

Our works, which I shall review now, were concerned with the last
possibility mentioned above, namely that $|1-q|$ measures fluctuations in
the hadronizing system under consideration. As shown in \cite{WW} in the
case when these fluctuations can be described by gamma distribution,
\begin{eqnarray}
  && \left[ 1 -(1-q) \frac{\varepsilon}{\lambda}\right]^{1/((1-q)} =
  \int_0^{\infty}  e^{-\frac{\varepsilon}{\lambda}}f(\beta)d\beta, \label{gamma}\\
  &&   f(\beta) = \frac{1}{\Gamma(\alpha)}
   \left(\frac{\alpha}{\langle \beta \rangle}\right)^{\alpha}
    \beta^{\alpha-1}e^{-\frac{\alpha}{\langle \beta \rangle}\beta}, \quad \alpha
    =\frac{1}{q-1};\label{f}\\
    && q = 1 \pm \frac{\langle \beta^2\rangle - \langle \beta
    \rangle^2}{\langle \beta\rangle^2} , \label{q}
\end{eqnarray}
where $\langle \dots \rangle $ are the respective averages taken with
respect to $f$. In general case one refers to the concept of the so
called {\it superstatistics} discussed at length in \cite{BC}. This
approach has been recently further generalized in \cite{Birofluct}.

The situation encountered can be visualized in the following way. Already
in \cite{VH} it was suggested that the overwhelmingly success of
thermodynamical models has its origin in the fact that out of the large
number $N$ of secondaries produced in a given event only one (very rarely
two) is chosen for making distributions. Inevitably, the remain $(N-1)$
act as a  kind of {\it heat bath} which action can in most cases be
described by single parameter $T$ (called usually {\it temperature}
because of associations with thermal models). In this case one has simple
exponential distributions (i.e., $q=1$). However, because such "heat
bath" is in reality neither homogeneous nor infinite, it is natural to
expect that it should be described by more parameters. The simplest
extension is to assume that $T$ fluctuates, replace $T \rightarrow
T_0=\langle T\rangle$ and introduce another (second) parameter describing
its fluctuations - this leads to $q\neq 1$ in most natural way.
Mathematically it can be realized by introducing to the well-known linear
Langevin equation with additive noise, modelling the Brownian motion of
some test particle (and leading to a Boltzmann distribution as stationary
solution of the corresponding Fokker-Planck (FP) equation), a small
multiplicative noise. This leads immediately to Tsallis distributions of
relevant observable as {\it exact} stationary solution of the
corresponding FP equation with $(q-1)$ being directly connected with the
strength of this new term (cf. \cite{WW,Birofluct}). To close this point
it must be mentioned that the role of correlations is so far not excluded
but it seems that in the realm of high energy multiparticle production
they are, at least at circumstances considered so far, not the dominant
component \cite{WWc,BGGM}.

Experimental data of interest come usually in two categories: as
distributions in longitudinal phase space (usually in rapidity $y$)
averaged over transverse momenta, $dN/dy$, or distributions in transverse
momenta, $dN/dp_T$, averaged over $y$ or taken for some window in $y$. We
have therefore also two kinds of "temperatures", $T_L$ and $T_T$, and
their fluctuations are thus described by, respectively, $q_L$ and $q_T$.
Out of these two classes only exponential shape of $dN/dp_T$ spectra is
usually regarded as indication of {\it thermal-like} character of such
processes with a kind of {\it local thermal equilibrium} setting in and
characterized by {\it temperature} $T=T_T$, which can be deduced from the
corresponding slopes of $dN/dp_T$. Any deviation from them is then
interpreted as signal for some {\it dynamical} (nonequilibrium) effects
showing up (like, for example, the flow or decay of resonances, see
\cite{FloRes} and references therein). Instead of trying to exclude them
one can investigate the possibility that the observed nonexponential
spectra result from some new form of equilibrium characteristic of
nonextensive thermodynamics (incorporating, among others, also effects of
the aforementioned factors).
\begin{figure}[!htb] \vspace*{-0.cm}
\begin{center}
\includegraphics*[width=3.5cm]{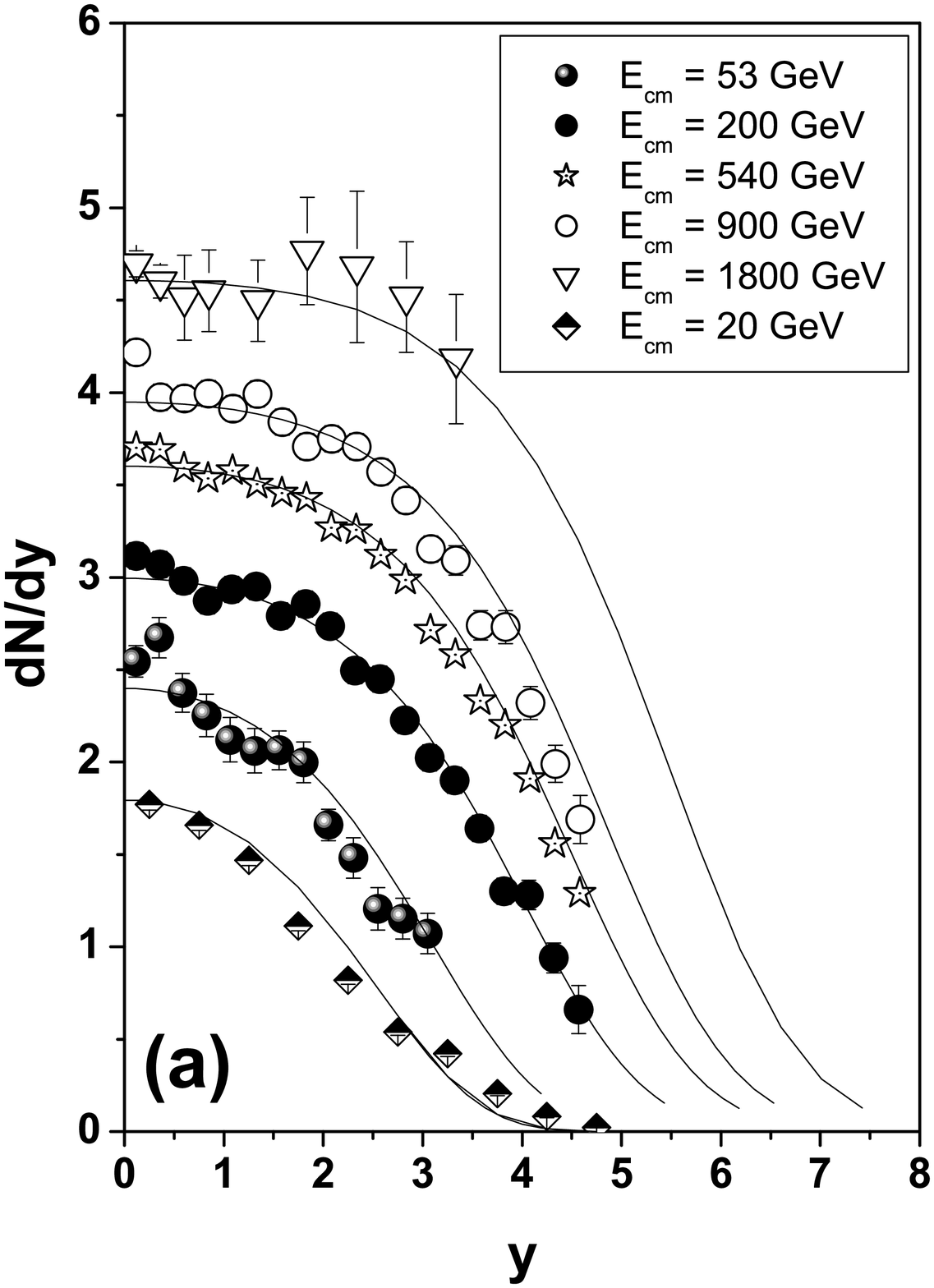}\hfill
\includegraphics*[width=3.34cm]{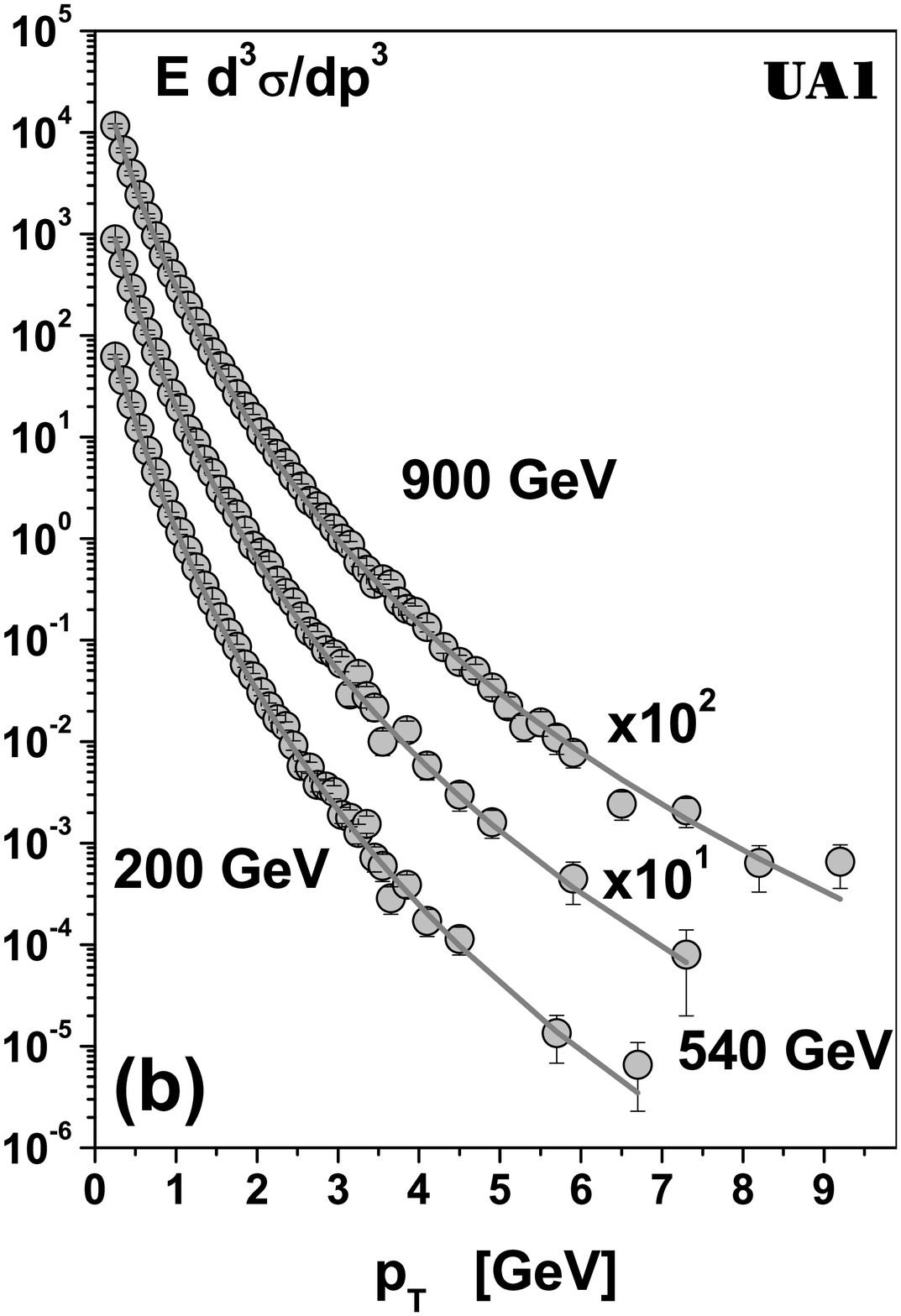}
\end{center}
\vspace*{-0.7cm} \caption{Examples of Tsallis distributions: $(a)$ fit to
rapidity spectra for charged pions produced in $pp$ and $\bar{p}p$
collisions at different energies. $(b)$ fit to $p_T$ spectra from UA1
experiment (see \cite{NextT} for details and references to experimental
data).} \label{f1-2}
\end{figure}
The plausibility of such approach is illustrated in Fig. \ref{f1-2} where
examples of both types of distributions are shown \cite{NextT}. The
characteristic feature encountered here is that $q_L>q_T$ and that $T_L
>> T_T$ ($q_L$ ranges in linear fashion from $1.05$ for $\sqrt{s}=20$ GeV
to $1.33$ for $\sqrt{s} = 1800$ GeV whereas variation of $q_T$ is limited
to between $1.095$ for $\sqrt{s} = 200$ GeV and $1.11$ for $\sqrt{s} =
900$ GeV; the respective changes of $T_L$ are from $1.76$ GeV to $55.69$
GeV whereas those of $T_T$ are much smaller varying from $0.134$ GeV to
$0.14$ GeV). The immediate question arises: what is the meaning of $q_L$,
or, equivalently, what is the meaning of fluctuations of "partition
temperature" $T_L$? The answer is that $ q_L = 1 + 1/k$, where $k$ is
parameter defining (in addition to mean multiplicity $\langle n\rangle $)
the Negative Binomial distributions of multiplicity observed in such
collisions \cite{MaxEntq}. It is because $T_L\sim M/\langle n\rangle$,
where $M$ is energy available for production of secondaries, therefore
for $M$ kept constant fluctuations of $T_L$ mean automatically
fluctuations of $\langle n\rangle$ and those lead immediately to NBD
\cite{MaxEntq}. Interestingly enough NBD arises also in a natural way
when particles are produced with energies distributed according to
Tsallis distribution (when they are distributed according to Boltzmann
distributions one gets Poisson distribution instead, see \cite{WWc}).

Let us proceed now to recent RHIC data on $p_T$ distributions, which
consist our potential source of the parameter $T$, as mentioned above. As
shown in \cite{MB1}, using approach based either on nonextensive
statistics or on stochastic ideas one can successfully account for the
{\it whole range} of the observed transverse momenta. This is because in
both cases the resultant distributions are {\it intrinsically
non-exponential}. In \cite{Hq} the same data were analyzed using Hagedorn
model with spectrum of resonances given by $\rho(m)$:
\begin{eqnarray}
    \frac{d^3\sigma}{dp^3} &=& C \int dm \rho(m)
        \exp\left(- \beta_0\cdot\sqrt{p_l^2 + p_t^2 + m^2}\right),
        \label{eq1}\\
\rho(m) &=& \frac{\exp(m\beta_H)}{(m^2 + m_0^2)^{5/4}}.  \label{eq2}
\end{eqnarray}
\begin{figure}[!htb]
\vspace*{-.2cm}
\begin{center}
\includegraphics*[width=4.33cm]{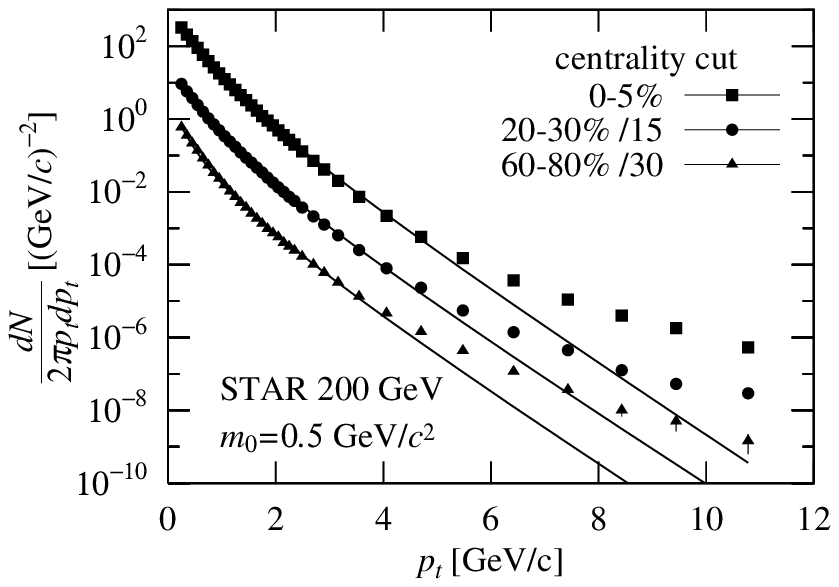}
\includegraphics*[width=4.24cm]{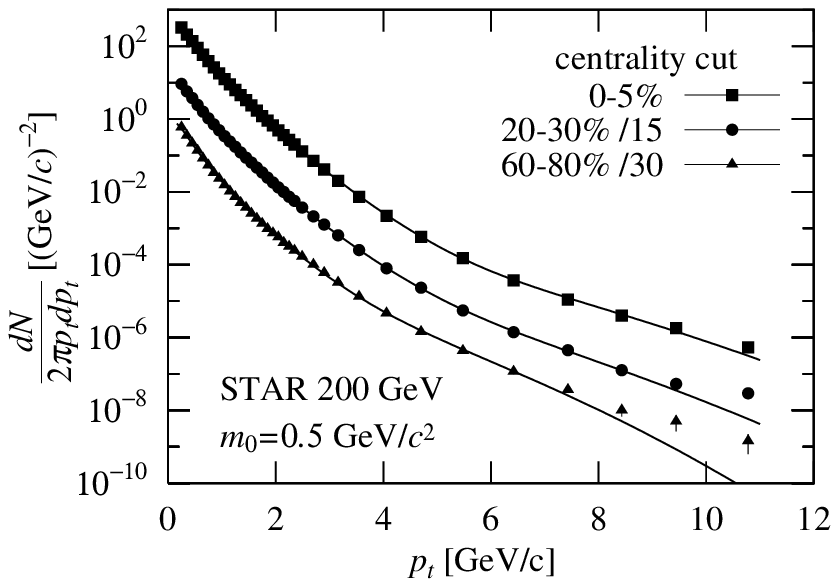}
\end{center}
\vspace*{-0.7cm} \caption{Example of analysis of STAR data~\cite{STAR} by
using usual Hagedorn formula (eq. (\ref{eq1}) with $q=1$, left panel) and
its nonextensive generalization (eq. (\ref{eq1}) with $q > 1$, right
panel), see \cite{Hq} for details.} \label{f_Hq}
\end{figure}
As seen in Fig. \ref{f_Hq}, although $\rho(m)$ introduces already some
fluctuations to the system, experimental data can be fitted {\it only}
with $q_T>1$ (of the order of $q_T\simeq 1.00015$). The new and
potentially very interesting fact is that similarly good fit can be
obtained with simple $q$-exponential (i.e., by putting $\rho(m)=1$ in eq.
(\ref{eq1})), however, in this case $q$ is noticeably greater and equal
to $q_T=1.065$. This effect is seen for all RHIC data and for all
centralities. It strongly suggests that including some well identified
sources of fluctuations to simple statistical model (represented here by
function $\rho(m)$ as given by eq. (\ref{eq2})) accounts for some
fluctuations and lowers therefore the value of parameter $q_T$. However,
as seen in Fig. \ref{f_Hq}, data are very sensitive to $q$ and, as has
been shown in \cite{Hq}, to obtain good estimation of $T_0=1/\beta_0$ one
has to resort to nonextensive version of eq. (\ref{eq1}) with $q_T>1$.
\begin{figure}[!htb]
\vspace*{0.1cm}
\begin{center}
\includegraphics*[width=6cm]{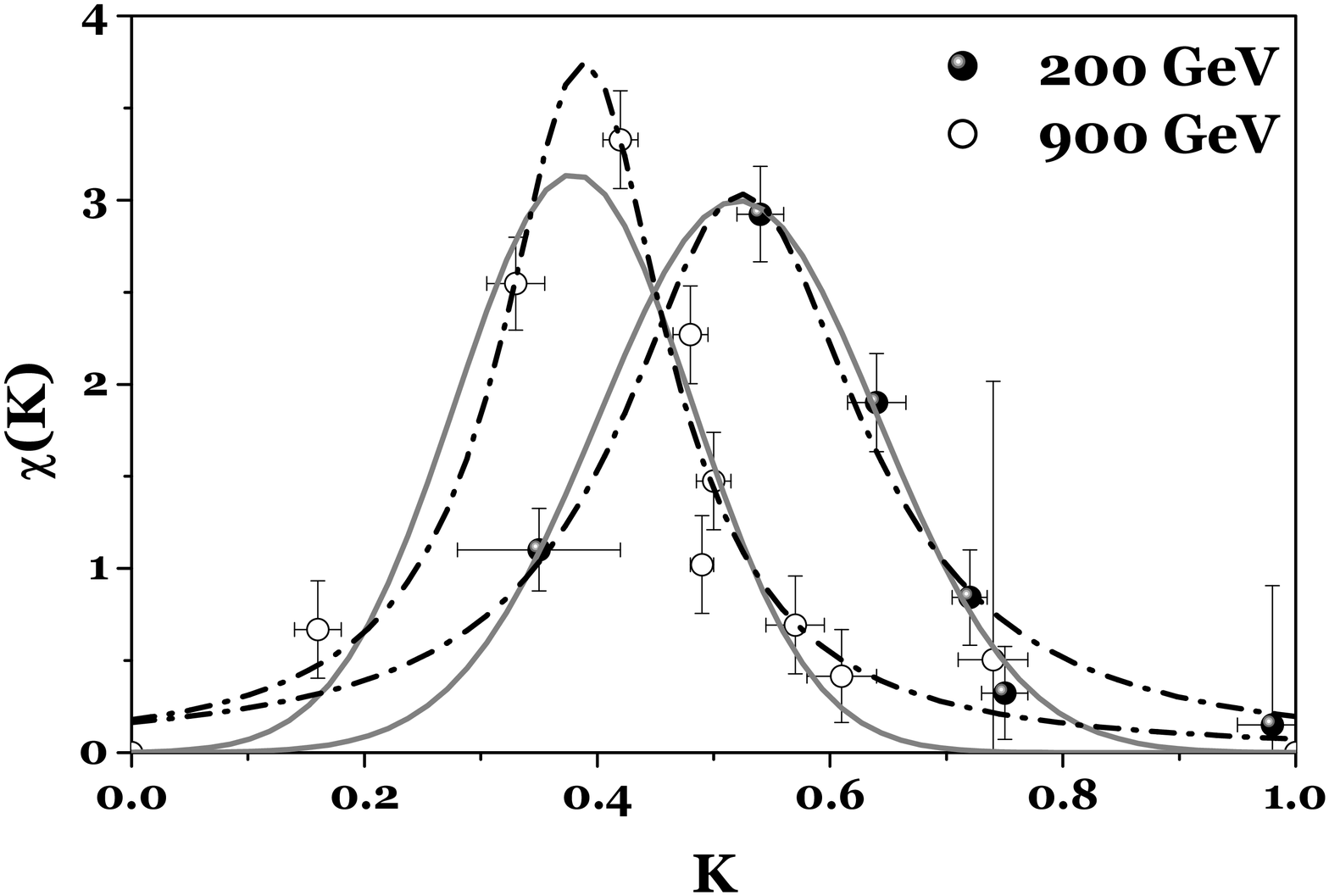}
\end{center}
\vspace*{-.6cm} \caption{Example of EGE \cite{TO}: Inelasticity
distributions $\chi(K)$ (normalized to unity) obtained in \cite{MaxEntq}
from analysis of multiparticle production data for $\sqrt{s} = 200$ GeV
and $\sqrt{s} = 900$ GeV fitted by gaussian (full lines) and lorentzian
(dashed lines) distributions, respectively (see \cite{MaxEntq} for more
details). } \label{f_khi}
\end{figure}

We would like to close this Section by mentioning that there is yet
another source of fluctuation in the production process, which also leads
to nonextensivity but not of Tsallis type, i.e., with gaussian
distributions instead of eq. (\ref{eq:defq}). Their source lies in the
fact that usually only fraction $K$ (called {\it inelasticity}) of the
total energy of reaction is used for production of secondaries and this
fraction can fluctuate. It turns out that such fluctuations are most
naturally described in the frame of the so called Extended Gaussian
Ensemble (EGE) by simple gaussian distribution, $\chi(K) \simeq \exp
\left[ -(K-\langle K\rangle)^2/(2\sigma^2) \right]$, (cf. \cite{TO} for
details and references) and that they can reasonably fit distributions of
$\chi(K)$ for different energies obtained in \cite{MaxEntq}, see Fig.
\ref{f_khi}. In fact, closer inspection of Fig. \ref{f_khi} shows that
still better fit can be obtained when using the so called lorentzian
distribution, which is nothing else as $q$-gaussian with $q=2$. It would
mean then that even here there are still some other fluctuations seen in
data, which are undisclosed by the usual EGE approach.

\section{Summary}
To summarize: we argue (since quite a time already) that whenever one
finds in data a hint for eq. (\ref{eq:defq}) the immediate suspicion
should be that they hide some intrinsic fluctuations with strength given
by $q-1$. What is the dynamical origin of these fluctuations is, however,
out of the scope of the procedure discussed here as $q$ accounts for {\it
all of them}. The $q>1$ results should then urge experimentalists to
devise some special measurements devoted to search for fluctuations.

To allow the reader to make personal judgement in what concerns the
possible role of correlations let us list recent attempts to interpret
parameter $q$ by some dynamical correlations caused either by incomplete
phase space occupance \cite{Others1} or by some specific changes
introduced in generalized form of the Boltzmann equation (either by using
corresponding collision rates nonlinear in the one-particle densities or
by using nontrivial energy composition rules in the energy conservation
constraint part,  cf. \cite{Others2} and references therein). It should
also be added at this point that, as advocated in \cite{Parvan}, one can
also view $q > 1$ as a general (leading order) finite-size effect.

Finally, one should be aware of the fact that there is still an ongoing
discussion on the meaning of the temperature in nonextensive systems.
However, the small values of parameter $q_T$ deduced from data allow us
to argue that, to first approximation, $T_0$ can be regarded as the
hadronizing temperature in such system. One must only remember that in
general what we study here is not so much the state of equilibrium but
rather some kind of stationary state (see \cite{Abe} and references
therein).

Let us close by saying that this subject is still an open issue for
further research (like, for example, event-by-event analysis of data
\cite{rest} or hydrodynamical models \cite{OW}).

\noindent{\bf Acknowledgements}

The subject reviewed here has been investigated in collaboration with
O.Utyuzh, Z.W\l odarczyk, F.S.Navarra, M.Biyajima, M.Kaneyama,
T.Mizoguchi, N.Nakamija, N.Suzuki and T.Osada. Partial support (GW) of
the Ministry of Science and Higher Education (grants Nr
621/E-78/SPB/CERN/P-03/DWM 52/2004-2006 and 1 P03B 022 30) is
acknowledged.

\end{document}